\documentclass[journal]{IEEEtran}
\usepackage[T1]{fontenc}
\usepackage{float}
\usepackage{amsmath}
\usepackage{amsthm}
\usepackage{amssymb}
\usepackage{stackrel}
\usepackage{graphicx}
\usepackage{setspace}
\usepackage{url}
\allowdisplaybreaks

\usepackage{caption}
\usepackage{subfigure}
\usepackage{algorithm}
\usepackage[graphicx]{realboxes}
\usepackage{algorithmic}
\usepackage{verbatim}

\makeatletter

\floatstyle{ruled}
\newfloat{algorithm}{tbp}{loa}
\providecommand{\algorithmname}{Algorithm}
\floatname{algorithm}{\protect\algorithmname}
\theoremstyle{plain}

\theoremstyle{definition}

\theoremstyle{plain}

\theoremstyle{plain}
\newcommand{\RNum}[1]{\uppercase\expandafter{\romannumeral #1\relax}}
\usepackage[thicklines]{cancel}
\IEEEoverridecommandlockouts
\usepackage{ifpdf}
\usepackage{cite}

\ifCLASSOPTIONcompsoc
\usepackage[caption=false,font=normalsize,labelfont=sf,textfont=sf]{subfig}
\else
\usepackage[caption=false,font=footnotesize]{subfig}
\fi
\usepackage{graphicx}
\usepackage{epstopdf}
\usepackage{mathtools}
\usepackage{multirow}
\usepackage{booktabs}

\usepackage{tabularray}

\usepackage{setspace}
\usepackage{lettrine} 
\usepackage{bm}
\makeatother
\usepackage{babel}

\usepackage{booktabs} 
\usepackage{longtable}

\usepackage{stfloats}
\usepackage{colortbl}  
\usepackage{xcolor}
\usepackage{array}   

\begin{document}
	\captionsetup[figure]{font={small}, name={Fig.}, labelsep=period}
	\title{
 Advancing Ubiquitous Wireless Connectivity through Channel Twinning}
	\author{
		Yashuai Cao,
        Linglong Dai,~\IEEEmembership{Fellow,~IEEE},
		Jingbo Tan,
		Jintao Wang,~\IEEEmembership{Senior Member,~IEEE},\\
        Tianyue Zheng,
		Wei Ni,~\IEEEmembership{Fellow,~IEEE},
		Ekram Hossain,~\IEEEmembership{Fellow,~IEEE},
		and Dusit Niyato,~\IEEEmembership{Fellow,~IEEE}
        \thanks{This work of Y. Cao was supported in part by the Project funded by China Postdoctoral Science Foundation under Grant No. 2023M742009; in part by the Postdoctoral Fellowship Program of CPSF under Grant No. GZC20231372; in part by the National Natural Science Foundation of China under Grant No. 62401315;
        in part by the National Research Foundation, Singapore, and Infocomm Media Development Authority under its Future Communications Research \& Development Programme, Defence Science Organisation (DSO) National Laboratories under the AI Singapore Programme (FCP-NTU-RG-2022-010 and FCP-ASTAR-TG-2022-003), Singapore Ministry of Education (MOE) Tier 1 (RG87/22), the NTU Centre for Computational Technologies in Finance (NTU-CCTF), and Seitee Pte Ltd.}
		\thanks{Y. Cao, L. Dai, J. Tan, J. Wang, and T. Zheng are with the Department of Electronic Engineering, Tsinghua University, Beijing 100084, China, and also with the Beijing National Research Center for Information Science and Technology (BNRist), Beijing 100084, China (e-mail: \{caoys, daill, tanjingbo, wangjintao\}@tsinghua.edu.cn; zhengty22@mails.tsinghua.edu.cn).}
		\thanks{W. Ni is with CSIRO, Sydney, New South Wales, 2122, Australia (e-mail: wei.ni@data61.csiro.au).}
		\thanks{E. Hossain is with the Department of Electrical and Computer Engineering, University of Manitoba, Canada (email: ekram.hossain@umanitoba.ca).}
		\thanks{D. Niyato is with the School of Computer Science and Engineering, Nanyang Technological University, Singapore 639798 (e-mail: dniyato@ntu.edu.sg).}
	}
	
	\maketitle
	\begin{abstract}
        As an emerging trend in channel acquisition (CA), the concept of channel twinning (CT) has been proposed as a powerful enabler of ubiquitous connectivity in next-generation (xG) wireless systems. By fusing multimodal sensor data, CT advocates a high-fidelity and low-overhead CA paradigm, which is promising to provide accurate channel prediction in cross-domain and high-mobility scenarios of ubiquitous xG networks. However, existing literature lacks a universal CT architecture to address the challenges of heterogeneous scenarios, data, and resources in xG networks, which hinders the widespread deployment and applications of CT. 
		This article discusses a new modularized CT architecture to bridge scene recognition, cooperative sensing, and decentralized training, comprising  versatile model configuration, multimodal cooperative sensing, and lightweight twin modeling modules.
		Additionally, this article presents a detailed concept, technical features, and case studies of CT, outlines mainstream trends of realization methods, followed by potential applications of CT-empowered ubiquitous connectivity, and issues requiring future investigations.
	\end{abstract}
	
	\begin{IEEEkeywords}
		Next-generation (xG) wireless communication system, channel twinning (CT), ubiquitous connectivity, multimodal sensing, lightweight digital twin. 
	\end{IEEEkeywords}
	
	\section{Introduction}\label{sec:1}
	\IEEEPARstart{U}{biquitous} connectivity is regarded as an overarching and fundamental function of next-generation (xG) networks (e.g., 6G networks), as it underpins seamless wireless coverage and network resilience to natural disasters~\cite{9786750}. 
	To have ubiquitous connectivity, advanced coverage extension solutions such as space-air-ground-sea-integrated network (SAGSIN), cell-free network, and reconfigurable intelligent surface (RIS) have been proposed~\cite{9373011}.
	The provisioning of zero coverage-hole networks has to deal with challenges related to diversified wireless channels across heterogeneous networks and multiple frequency bands, thus rendering network planning and optimization more convoluted than ever.
	This necessitates attention to one of the fundamental challenges in wireless communication systems: channel acquisition (CA), which determines the performance limits of wireless networks.

	\subsection{Advances in Channel Acquisition}\label{sec:1a}
	Recently, channel twinning (CT) has emerged as an innovative approach to CA that captures the intricate interactions between channels and physical environments. CT has received increasing attention for its ability to enhance traditional CA in the following ways:
    \subsubsection{From mathematical to environmental modeling}
    It can be difficult for mathematical models (statistical and deterministic models) to strike a balance between channel prediction accuracy and computational complexity~\cite{10198575}. Instead, CT aims to reveal the channel mapping from environmental attributes and achieve an accurate CA with low complexity.
    \subsubsection{From unimodal to multimodal sensing}
    Traditional CA primarily relies on radio frequency (RF) sensing-based pilot training methods, accompanied by high overhead. In contrast, multimodal fusion~\cite{9662100, 10330577} is a pillar of CT, and it is promising to reduce or even eliminate RF pilots.
    \subsubsection{From physical-world to enhanced digital-world processing}
    Traditional channel modeling or channel estimation methods often rely on channel assumptions to fit the channel model solely based on measurements. CT exploits measurements to extract the environment semantics or interpolate the channel measurements at any location, leading to a channel knowledge model that better reflects real-world environments. On the other hand, CT has the potential to significantly reduce the dependency on expensive real-world measurements once the CT model is constructed.

	Table~\ref{tb:1} provides a conceptual comparison to clarify the relationship between CT and other related concepts.
\begin{table*}[h!]
\centering
\fontsize{8pt}{0.6\baselineskip}\selectfont
\caption{\small{Comparison between different concepts about channel acquisition.}}
\begin{tblr}{
  width = \linewidth,
  colspec = {Q[120]Q[140]Q[160]Q[160]Q[120]Q[180]},
  cell{1}{5} = {c=2}{0.3\linewidth},
  cell{2}{1} = {r=3}{},
  cell{2}{5} = {c=2,r=3}{0.3\linewidth},
  cell{4}{3} = {c=2}{0.3\linewidth},
  cell{5}{5} = {c=2}{0.3\linewidth},
  cell{6}{5} = {c=2}{0.3\linewidth},
  cell{7}{1} = {r=3}{},
  cell{7}{5} = {r=3}{},
  vlines,
  hline{1-2,5-7,10} = {-}{},
  hline{3-4} = {2-4}{},
  hline{8-9} = {2-4,6}{},
}
\textbf{Concept} & \textbf{Modeling Method} & \textbf{Pros} & \textbf{Cons} & \textbf{Role and Application} & \\
{\textbf{Chanel }\\\textbf{modeling}}   
& Statistical & Versatility & Limited accuracy & 
{\labelitemi\hspace{\dimexpr\labelsep+0.5\tabcolsep}\textbf{Role}: Derive effective model representation\\\labelitemi\hspace{\dimexpr\labelsep+0.5\tabcolsep}\textbf{Application}: Pre-deployment}  & \\
& Deterministic & High accuracy & Time-consuming  &  &  \\
& Hybrid  & Balance accuracy and complexity  &  &  &  \\
{\textbf{Channel}\\\textbf{estimation}} & Pilot-based & Stable performance                                                            & {\labelitemi\hspace{\dimexpr\labelsep+0.5\tabcolsep}High pilot load\\
\labelitemi\hspace{\dimexpr\labelsep+0.5\tabcolsep}Outdated CSI} & {\labelitemi\hspace{\dimexpr\labelsep+0.5\tabcolsep}\textbf{Role}: Obtain current CSI\\
\labelitemi\hspace{\dimexpr\labelsep+0.5\tabcolsep}\textbf{Application}: Post-deployment}  & \\
{\textbf{Channel}\\\textbf{prediction}} & Time series analysis     & Proactive management & Error accumulation  & {\labelitemi\hspace{\dimexpr\labelsep+0.5\tabcolsep}\textbf{Role}: Infer future channel states\\\labelitemi\hspace{\dimexpr\labelsep+0.5\tabcolsep}\textbf{Application}: Post-deployment}   &  \\
{\textbf{Channel}\\
\textbf{twinning}}   
& Spatial interpolation    
& {\labelitemi\hspace{\dimexpr\labelsep+0.5\tabcolsep}Simple implementation\\
\labelitemi\hspace{\dimexpr\labelsep+0.5\tabcolsep}Low data dependency}   & {\labelitemi\hspace{\dimexpr\labelsep+0.5\tabcolsep}Limited accuracy\\
\labelitemi\hspace{\dimexpr\labelsep+0.5\tabcolsep}Poor scalability}  & {\labelitemi\hspace{\dimexpr\labelsep+0.5\tabcolsep}\textbf{Role}: Create the digital replica of the channel evolution\\
\labelitemi\hspace{\dimexpr\labelsep+0.5\tabcolsep}\textbf{Application}: Full life cycle} & Stable scenarios with smooth channel changes \\
& Environment semantics     & Enhanced modeling accuracy   & Expensive environment data collection                             &  & Scenarios with significant environment structure effects \\
& Deep learning   & 
{\labelitemi\hspace{\dimexpr\labelsep+0.5\tabcolsep}Automatic learning \\
\labelitemi\hspace{\dimexpr\labelsep+0.5\tabcolsep}No explicit assumptions} & {\labelitemi\hspace{\dimexpr\labelsep+0.5\tabcolsep}Need large data \\
\labelitemi\hspace{\dimexpr\labelsep+0.5\tabcolsep}Hyper-parameter tuning} &   & Highly dynamic and complex scenarios
\end{tblr}
\label{tb:1}
\end{table*}
 
	\subsection{Current Status of CT Studies}\label{sec:1b}
	In the context of CT, some initial studies have been reported in~\cite{9540134, 9897088, wang2023towards}.
	The term of digital twin channel (DTC)\footnote{To emphasize that ``channel twinning'' is a technology, similar to channel modeling or channel sounding, we have not used the terminology ``DTC''.} appeared earlier in~\cite{9540134}, but the concept has not been specified. 
    Later, the potential of channel recreation on digital twin (DT) was revealed in~\cite{9897088}, although the DTC concept was yet to be proposed.
    More recently, DTC has been defined as the mapping between channels and associated communication operations~\cite{wang2023towards}. Therein, the radio environment knowledge pool (REKP) was proposed to support DTC.

	The above literature has focused primarily on channel knowledge map (CKM) applications~\cite{9373011} or only discusses the REKP construction from a local and centralized perspective. While providing a framework for reflecting propagation environment via channel fingerprints, CKM emphasizes static mappings. REKP~\cite{wang2023towards} is a knowledge semantic database that stores and manages the feature relationships of radio environments and channels. 
	However, it is not tailored to specific scenarios or environments. In xG networks, deploying CT faces the challenges of cross-domain modeling, distributed sensing, and lightweight training. A centralized strategy makes it hard to cater to heterogeneous resources.
 
    In this article, a systematic discussion on the modularized architecture of ubiquitous CT is conducted to facilitate the organic integration of CT and ubiquitous sensing and intelligence. 
	For a better understanding of CT, a concise definition and several clarifications regarding the general CT misconceptions are refined. 
	This article also highlights the technical evolutions, the potential use cases for CT-empowered ubiquitous connectivity technologies, and several promising directions to spur future research.

	\section{Basic Concepts and Features of CT}\label{sec:2}
	\subsection{Concepts and Realization Principle of CT}\label{sec:2a}
\begin{figure*}[t]
		\centering{}\includegraphics[width=5.8in]{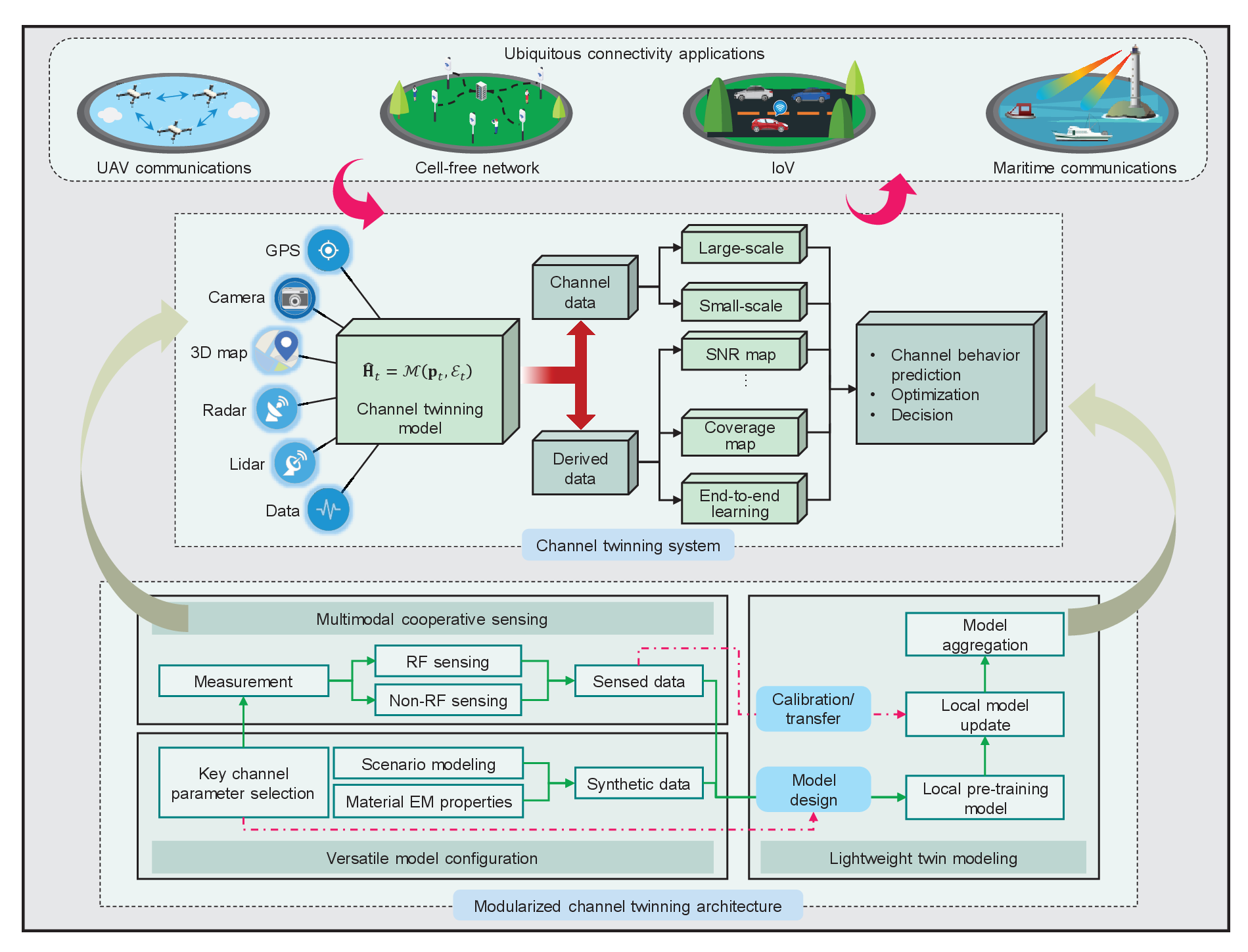}
		\caption{Conceptual framework of CT in xG networks and its modularized CT architecture. The versatile model configuration module provides information for measurements and model design to the other two modules. It generates synthetic data to pretrain the CT model in the lightweight twin module. Then, the CT model can be fine-tuned by feeding data from the multimodal cooperative sensing module.}
		\label{fig:1}
	\end{figure*}

	CT is defined as a digital mirror construction of physical channels for network design, optimization, and evaluation without real deployment. 
    In essence, such a digital mirror is a mapping model that indicates the correspondence from the environment to various aspects of associated channel behaviors, such as channel state information (CSI), channel path angle, or channel gain, depending on the specific task.
    Specifically, a site-specific CT aims to simulate the RF propagation law. 
	As shown in Fig.~\ref{fig:1}, CT can acquire CSI at any position based on the environmental conditions.

    Current methods for realizing CT fall into three main categories: Spatial interpolation (SI) modeling, environment semantics (ES) modeling, and deep learning (DL) modeling.
    \subsubsection{SI}
    SI leverages the spatial correlations of sampling channel measurements within the region of interest (RoI) to predict the channel of unobserved locations. Classic SI methods include inverse distance weighting, Kriging, etc., commonly used for constructing CKM~\cite{9373011}.
    \subsubsection{ES}
    Utilizing ES features (e.g., building layout, terrain type, and obstacles), ES modeling can help accurately build channel models, such as the virtual obstacle model~\cite{10041012} and REKP~\cite{wang2023towards}.
    \subsubsection{DL}
    DL methods apply advanced deep networks to learn the complex mapping between various environmental factors affecting the channels. In DL methods, differentiable ray tracing (DRT) is gaining popularity as a promising DT enabler~\cite{10465179}. The DRT principle is to learn the electromagnetic (EM) properties of involved materials by modeling the interaction between ray paths and the environment.
 
	In practice, suitable methods should be selected according to applications and dataset sizes. 	
	From the perspective of implementation complexity, CT deployment should undergo three evolution stages: i) Synthetic CT~\cite{9897088}, which is deployed in a quasi-static environment to generate synthetic data that approximates the measurements in the RoI; ii) Transfer CT~\cite{9771088}, which constructs self-adaptive CT based on knowledge transfer to cope with new scenarios; and iii) Multiverse CT~\cite{10478322}, which customizes different twins to offer multiple fidelity levels for applications with diverse computing power and delay constraints.

	\subsection{Common Misconceptions and Four Clarifications}\label{sec:2c}
	To elaborate on the concept of CT, we discuss the CT model illustrated in Fig.~\ref{fig:1} from inputs to outputs. Several basic features are highlighted as follows: 
    a) \textbf{context-awareness}: Instead of merely using received pilots or feedback, CT integrates contextual information from multimodal sensors (e.g., camera, GPS, and Radar) to learn environmental dynamics; 
    b) \textbf{high-fidelity}: Thanks to three-dimensional (3D) map-aided methods~\cite{10041012, 10465179}, e.g., ray-tracing, CT allows for high-fidelity emulation of complex interactions between the environment and the network; 
    c) \textbf{self-synchronization}: CT can model the whole life cycle of channel behaviors in the environment for real-time emulation; 
	d) \textbf{expertise-driven}: EM and channel modeling expertise help design compact DL structures and improve interpretability when specific channel properties, e.g., signal-to-noise ratio (SNR), are desired to be output; and
	e) \textbf{region-specific}: Since the CT model is trained on measurements from a specific site's coverage region, the DTC must be region-dependent.
    

	Since CT is an emerging technology, several common misconceptions need to be clarified. 
	
	\vspace{2 mm}
    \textbf{Clarification 1:} \emph{Do statistical and deterministic modeling belong to CT technology?}
	\vspace{2 mm}
	
    CT is designed to create a digital mirror of the channel response in a given environment, whereas standard statistical models like Rayleigh and Rician cannot match specific environments. Besides, 3GPP urban microcell (UMi) and urban macrocell (UMa) models only consider macroscopic environmental attributes (e.g., urban or rural) but ignore the specific environmental factors~\cite{10198575}. In this sense, statistical models may not reflect the CT features. 
    Deterministic models, such as ray tracing, is a physics-based model that predicts channel by simulating ray propagation in specific environments and scene geometries, as well as complex EM interactions with surrounding objects.
	On the other hand, CT is a mapping model-based CA method that seeks to efficiently mimic channel behaviors with no need for exhaustive recursive tracing. As a preliminary attempt at CT, Sionna is ten times faster than traditional ray-tracing like Wireless InSite (WI)~\cite{10465179}.
	Conventional channel modeling methods lay a solid foundation for understanding the physical behaviors of waves, while CT can adaptively respond to environmental changes through its self-synchronization capability, as discussed in Section II-B.
	Combining ray tracing with CT can further contribute to advanced channel modeling, and potentially bring real-time ray tracing to reality.

	\vspace{2 mm}
	\noindent\textbf{Clarification 2:} \emph{Does CT necessarily mean the use of DL?}
	\vspace{2 mm}

    According to specific application requirements, one can flexibly choose different CT training methods, e.g., SI, ES, and DL modeling methods.
	
	\vspace{2 mm}
	\noindent\textbf{Clarification 3:} \emph{Does using only position information to predict network operations belong to CT technology?}
	\vspace{2 mm}
	
	CT is either explicitly or implicitly context-aware. Only position-based methods construct the channels under the simple line-of-sight (LoS) assumption. However, blockages and scatters may cause non-LoS (NLoS) cases. This shows the significance of context awareness, and that position-based channel recovery cannot compete with CT.

    \vspace{2 mm}
    \textbf{Clarification 4:} \emph{What are the conceptual differences between DT, CT, and DTC?}
	\vspace{2 mm}

	A DT is a virtual replica of a physical entity (e.g., a network or device). 
	DTC develops an REKP to help understand the interaction between channels and environmental characteristics across extensive scenarios~\cite{wang2023towards}. 
	The REKP framework updates the path loss feature relationship graphs for different receiver locations through real-time acquisition of environmental features (e.g., position, blockage, and distance). However, these updates rely on static environmental factors within localized time intervals, rather than comprehensive dynamic modeling of channel evolution over time. 
	In contrast, CT focuses on learning both static attributes and the temporal evolution of channels, emphasizing the twinning process in a specific region.

	\section{A Modularized Design of CT Architecture}\label{sec:3}
	For the design and deployment of CT in xG networks, the following questions need to be answered.
	\begin{itemize}
		\item[\textbf{Q1}.] Since xG networks are anticipated to involve massive cross-domain links across space/frequency/time domains, how can we streamline the CT model deployment and configuration process for diverse scenarios and channel types?
		\item[\textbf{Q2}.] How do we effectively integrate multimodal sensor networks during the CT deployment?
		\item[\textbf{Q3}.] How can CT training in network nodes with limited computing capacity and power supply be realized?  
	\end{itemize}
	In light of these questions, we present a modularized architecture to discuss the design considerations for CT deployment in xG networks. The overall CT architecture comprises three modules of versatile model configuration, multimodal cooperative sensing, and lightweight decentralized CT training, as shown in Fig.~\ref{fig:1}.

	\subsection{Versatile Model Configuration Module}\label{sec:3a}	
	Various scenarios in xG networks give rise to various types of channels. The characteristics of these channels differ significantly from each other. 
	Consequently, a major challenge that arises for CT deployment is how to integrate the diverse channel characteristics across scenarios to facilitate the automated construction of suitable CT models. To this end, a versatile model configuration module tailored for ubiquitous CT is needed.

    Generally, a channel consists of large-scale fading (LSF) and small-scale fading (SSF) components, each with different characteristics in various scenarios. Therefore, the channel components associated with the scenario and working band should be identified first. For instance, high-mobility scenarios, such as Internet of Vehicles (IoV) and unmanned aerial vehicle (UAV) networks, should consider non-stationary channels. Gas absorption and blocking effects should be considered in millimeter-wave (mmWave) and Terahertz (THz) networks, while visible light communication (VLC) channels only involve LSF. Extremely large-scale, multiple-input-multiple-output (MIMO) channels should consider spherical wavefront and spatial non-stationarity~\cite{9786750}.
	
    To effectively deploy CT, we can quickly determine the channel components by following two steps: 1) LSF channel construction by a static 3D map; and 2) SSF channel prediction by capturing dynamic activities.
    The versatile model configuration module allows developers to easily configure the CT model by selecting from various model settings and adjusting parameters to suit specific task needs. Additionally, this module provides supplementary synthetic data to assist with CT model training, rather than engaging in traditional channel modeling.

	\subsection{Multimodal Cooperative Sensing Module}\label{sec:3b}
    With the rise of Internet of Things (IoT) devices and advances in integrated sensing and communication (ISAC), multimodal sensing has become increasingly common.
    Multimodal sensing data fusion (M\textsuperscript{2}SDF) is imperative for CT model training~\cite{10330577}. In the multimodal cooperative sensing module shown in Fig.~\ref{fig:1}, there are two main categories of multimodal sensors: RF and non-RF.

	The traditional channel measurements come primarily from RF reception and sensing. Such measurement is limited by the sensing range and accuracy of the single-site ISAC scheme. For RF sensing, a multi-site cooperative sensing scheme is preferable, which can enhance signal acquisition by ubiquitous connectivity. In this scheme, non-terrestrial and terrestrial base stations (BSs) can sense any position through ISAC. In addition, users can also track targets by the dedicated positioning reference signals defined in 5G-NR R17~\cite{10471311}. 
 
    However, cooperative sensing requires precise space-time registration, which is crucial for positioning data acquisition and continuous updates of CT systems. Non-RF sensing involves more heterogeneous multimodal data than RF sensing. In this sense, the challenge of M\textsuperscript{2}SDF lies in how to deal with the data heterogeneity at the CT server.
    ES plays a crucial role in enabling CT by providing detailed information about the physical environment. The ES implementation in the CT framework consists of the following steps: 1) Extracting physical structure features of the environment that impact signal paths, such as walls, trees, and other obstructions~\cite{10041012}; 2) classifying and encoding these environment features to construct channel semantic information. To extract propagation ESs, semantic alignment, and completion~\cite{wang2023towards} techniques may be invoked; and 3) utilizing semantic features to model the interactions between network actions and specific tasks. Thus, optimization algorithms or deep networks can be designed to improve the accuracy of CA, adapt to dynamic scenarios, or enhance communication performance.

	\begin{figure}[t]
		\centering{}\includegraphics[width=3.5in]{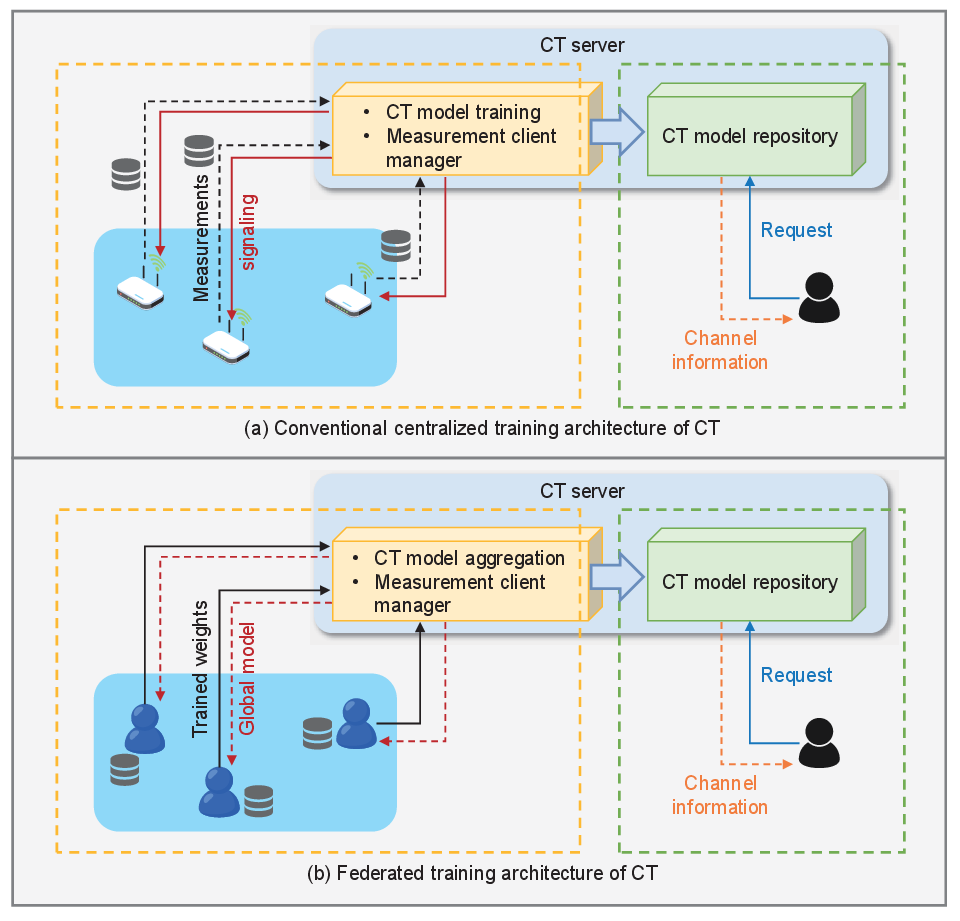}
		\caption{Training architecture of CT: (a) Centralized learning-based CT, where dedicated measurement anchors collect data and upload it to the CT server for training; and (b) FL-based CT, where registered device clients use their local data to train the model and only upload trained parameters for model aggregation at the CT server.}
		\label{fig:2}
	\end{figure}
 
	\subsection{Lightweight Twin Modeling Module}\label{sec:3c}
	Existing CT models typically train in a centralized manner. However, scalability and privacy challenges limit the CT deployment in ubiquitous connectivity. In this case, federated learning (FL) is viewed as an adequate and promising choice for lightweight CT because FL training is decentralized and does not require sharing data. 
    Fig.~\ref{fig:2} shows a lightweight CT training module based on FL. 
	In the FL-enabled CT scheme, FL clients only need to upload local model parameters to train a global CT model collaboratively, thus reducing communication costs and privacy leakage risks.
 
	The integration of FL and DT is not straightforward, as two practical issues need to be addressed as prerequisites. First, participating in FL tasks can increase the energy costs of client devices. Self-interested devices may be reluctant to participate in FL.
    Thus, proper FL incentive mechanisms~\cite{9743558} are needed to encourage participation while considering energy constraints, especially for energy-limited devices like UAVs.
    To achieve such energy-efficient FL incentives, we need to quantify the contribution of clients to the global CT model convergence and design suitable payment mechanisms. 
    Second, the heterogeneity of local data across clients requires handling data distribution imbalances, as personalized FL could become promising for lightweight CT in the future.

	\section{Case Studies: Intelligent Coverage and Lightweight Twin}\label{sec:4}
    This section provides two case studies to demonstrate the significance of CT for ubiquitous connectivity and discuss the lightweight FL-based CT performance.

    \begin{figure*}[ht!]
		\centering
        \includegraphics[width=6.5in]{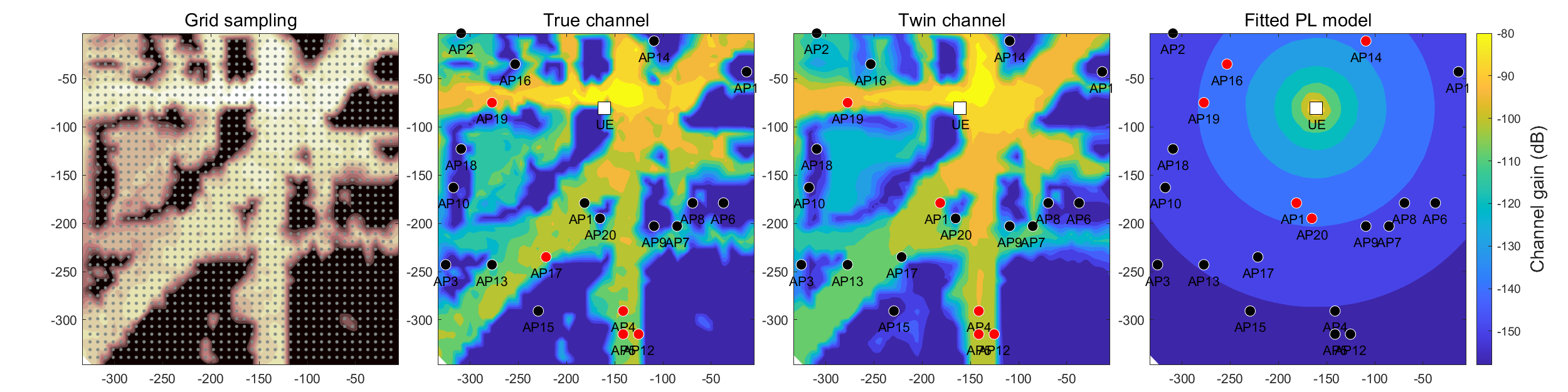}
		\caption{User association results produced based on the ground-truth channel, the twin channel, and the fitted PL model.}
		\label{fig:3}
	\end{figure*}

	\subsection{CT-Based Intelligent Coverage}\label{sec:4a}
	In ubiquitous connectivity networks, user association can dynamically adjust the connection of user equipment (UE) according to channel conditions to ensure continuous and efficient coverage.
	By selecting optimal access points (APs) for UE access, user association can improve resource utilization, reduce packet loss, and prevent interruptions. However, accurate full CSI is essential for effective user association.

	Cell-free communication is taken as an example to illustrate the benefits of CT for ubiquitous connectivity. In cell-free networks, obtaining the full CSI of involved nodes is challenged by limited pilots and high CSI exchange overhead.
	To address this, we propose a CT-based intelligent user association scheme. Specifically, the scheme is implemented via offline CT model training and online CSI query, as illustrated in Fig.~\ref{fig:2}(a). During the offline phase, dedicated measurement anchors collect channel measurements in the RoI and upload measurements to the CT server for training. Once trained, the CT model is stored in the twin repository. During the online phase, UEs request the CT server to query the CSI of the node they attempt to communicate with.

    We leverage the synthetic data generated by the WI software~\cite{wang2023towards} to train a CT model. As shown in Fig.~3, the transceiver locations are selected from grid points with a uniform spacing of 8 meters. We then use a deep neural network with seven hidden layers to capture how the environment affects the channel gain distribution. The network takes four-dimensional transceiver coordinates as input and outputs the corresponding channel gains. For any given transmitter location, we perform SI~\cite{9373011} using the predicted values from 30 randomly selected receiver positions, enabling the generation of the channel gain maps.

    Fig.~\ref{fig:3} shows channel gain maps with the UE marked by a white square. Conventional user association relies on the distance or path loss (PL) only. We fit a PL model with the same synthetic data by least squares fitting as a benchmark. By comparing the maps in Fig.~\ref{fig:3}, CT can efficiently perceive the environment, while the fitted PL model fails to reproduce the environment profile.
	
	With the predicted channel gain, we select five out of 20 distributed APs within the RoI for user association. The PL model can only select APs according to the distance criterion, as it does not grasp the ESs.
    By contrast, CT helps the UE avoid obstacles rather than selecting the nearest APs. 
    In addition, OpenStreetMap (OSM) data~\cite{10041012} can assist the CT model by generating synthetic data, particularly in scenarios where RF measurements might be sparse or unavailable, the CSI of any AP-UE link can be retrieved by CT.

 	\begin{figure}[t]
		\centering{}\includegraphics[width=2.5in]{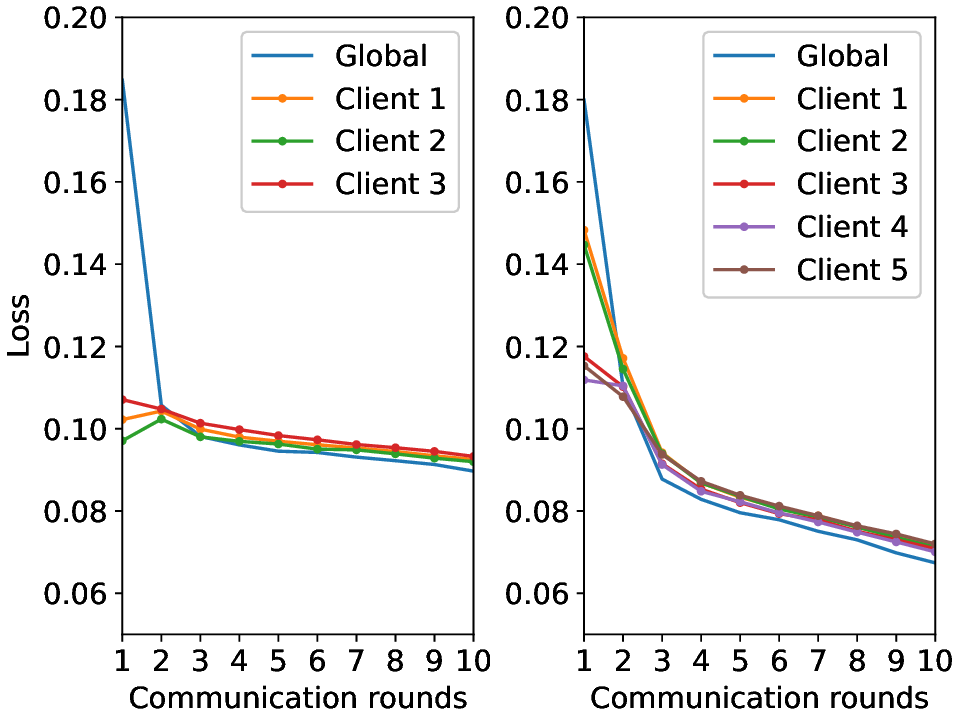}
		\caption{Validation loss of MSE versus communication rounds, where the total number of training samples is 10,000, and these samples are randomly divided between three or five UEs for FL.}
		\label{fig:4}
	\end{figure}
	
	\begin{figure}[t]
		\centering{}\includegraphics[width=2.5in]{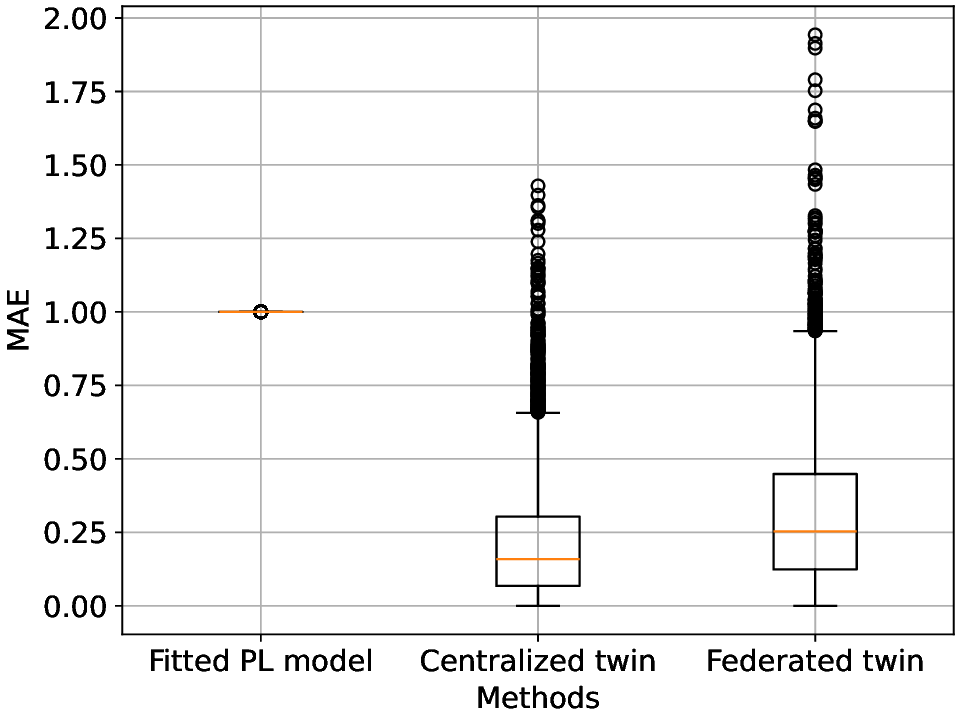}
		\caption{MAE comparison of channel gain prediction across different models.}
		\label{fig:5}
	\end{figure}
	
	\subsection{Federated Learning-Based Lightweight CT}\label{sec:4b}
    Centralized training of DT relies on sufficient computing capability and energy supply, which may be unavailable in sensor devices, e.g., UAVs. Fortunately, FL can realize a decentralized DT model construction while avoiding data privacy issues. In light of this, we investigate an FL-based lightweight channel gain twinning model; see Fig.~\ref{fig:2}(b).

	In the FL-based CT training architecture, no dedicated measurement anchors are needed. Mobile UEs act as potential measurement anchors. Incentive UEs can record their locations and associated channel measurements as they move into the RoI.
    Unlike centralized training, the UEs first register their identities with the CT server's client manager for subsequent FL training. When the CT server prepares to train a CT model, a group of registered UEs train their own CT models locally by using their recorded measurements. Then, they upload only the trained parameters to the server for model aggregation. The CT server updates the global model and sends model parameters back to the registered clients for local model updates. After several rounds of interactions, the global model converges, and hence, the FL-based CT training is completed.
	
	A portion of the channel gain dataset was randomly split among FL clients.
    Fig.~\ref{fig:4} shows the training loss of mean squared error (MSE). 
    It is seen that more clients contribute to lower MSE loss, even with the same total data samples. This is because more clients help smooth out the gradient noise of individual clients and improve the federated twin performance, though communication overhead rises.

    Fig.~\ref{fig:5} compares the channel gain (dB) prediction capabilities in terms of mean absolute error (MAE).
    The yellow solid line denotes the median MAE, the box represents the distribution range of the majority of MAE, and the black dots outside the box denote outliers. The PL model exhibits high prediction errors overall, indicating poor prediction accuracy.
    Meanwhile, few outliers indicate prediction stability, as the channel gain produced by the PL model lacks shadowing compared to the real channel.
    Both CT models show better prediction accuracy. Despite the fluctuations in prediction error of CT models, occasional outliers do not affect the overall performance of ubiquitous connectivity.


    In FL, the primary feedback overhead arises from the transmission of model updates from the UE to the BS. The feedback overhead of the 3GPP Type II codebook scales with the rank, and requires hundreds of bits (more than 500 bits for some system setups)~\cite{SHI2022413}. Since the overhead of different setups varies greatly, it is difficult to pre-determine the required overhead. Thus, the BS allocates physical uplink shared channel (PUSCH) resources for CSI reporting according to the maximum overhead, which causes resource waste and reduced efficiency. In this case, the UE can utilize leftover PUSCH resources for FL model parameter uploads after completing CSI reporting, thus improving resource utilization. Additionally, once the FL model is trained for large time-scale invariant environments, it will significantly reduce the channel feedback overhead.

	\section{Convergence of CT and Ubiquitous Communications: Opportunities and Challenges}\label{sec:5}
    In addition to reducing pilot overhead and enhancing prediction accuracy, CT can benefit ubiquitous connectivity by offering enhanced dynamic system robustness~\cite{10478322}, accelerated network development and testing~\cite{10198575}, and improved resource management~\cite{9662100}. To unfold the potential of CT, this section presents some open issues concerning the convergence of CT and ubiquitous communications. 
    
	\subsection{Opportunities of CT for Ubiquitous Connectivity}\label{sec:5a}
	\subsubsection{CT-Empowered Non-Terrestrial Communications}	
	Due to the high mobility of non-terrestrial systems, both satellite-to-ground and UAV-to-ground channels suffer from the non-negligible channel aging effect. Unlike the ground channel, UAV-to-ground channel gains do not show smooth changes with the movement of the UAVs. Due to the down-tilted antenna pattern, UAVs may easily fall into the antenna nulls of nearby BSs.
    Therefore, the UAVs cannot use the distance criterion to choose the cell. In this case, CT can emulate the antenna gain and airframe shadowing~\cite{9786750, 9662100} to achieve high-efficiency cell association in UAV networks.

	\subsubsection{CT-Empowered IoV Communications}
	One of the most immediate applications of CT is predictive beamforming in IoV. CT-enabled beamforming requires only necessary control signaling without extra pilots~\cite{10330577}.
	On the other hand, RIS is regarded as a low-power solution to dynamic blockage issues in IoV. However, CSI acquisition and frequent beam alignment in high-mobility scenarios hinder the widespread application of RIS. Through the CT model dedicated to serving the RIS sites, adaptive beam width control and vehicle tracking can be further realized.

	\subsubsection{CT-Empowered Maritime Communications}
    Since the maritime communications environment has distinct features of sparse scatters compared to terrestrial networks, CT is expected to enhance maritime connectivity by employing environmental information for network planning~\cite{10198575}. 
    Channel interference can be acquired in advance to avoid the acquisition burden of full CSI. 
    In the harsh underwater environment, underwater acoustic (UWA) channels suffer from spatio-temporal-frequency variability, limited bandwidth and severe noise and interference~\cite{9786750}. Channel estimation and data transmission become even more challenging in multi-node UWA communications. In this case, CT also brings unprecedented opportunities for UWA communications.

	\subsection{Challenges and Open Issues}\label{sec:5b}
	
	\subsubsection{Data Deficiency}
	CT demands sufficient high-quality measurement data, but channel measurement campaigns are costly and time-consuming. In practice, only sparse measurements from limited points are available. For example, DL-based indoor propagation prediction~\cite{9771088} only uses measurements from 36 locations. Furthermore, some measurements may contain errors due to hardware defects. To achieve high-precision CT, it is necessary to study the efficient utilization of synthetic and measurement data.

	\subsubsection{Personalized FL Mechanism of CT}
    Since FL clients only collect channel measurements from the local irregular areas, a simple federated averaging-based aggregation strategy can hardly be optimal. In particular, personalized FL with incentive mechanisms~\cite{9743558} needs to consider the CT model contribution of each client to the entire environment map.

	\subsubsection{Malicious CT Deployment}
	Malicious nodes with CT ability may exist in networks. They may access CSI at any location of interest. Therefore, once a malicious CT can recover CSI, wireless networks will face unparalleled security threats. Existing physical layer security and secret key schemes will be rendered ineffective.

    \subsubsection{High-Dimensional DTC Representation}
    While DTC is effective for wireless network optimization, high-dimensional channel matrices caused by large-scale MIMO or RIS raise computational and caching challenges for CT deployment. In this case, an equivalent channel compression representation~\cite{9897088} scheme in CT is attractive for network optimization.

	\section{Conclusion}\label{sec:6}
	From the perspective of the physical environment, a new CA paradigm, CT technology, has been studied to support xG ubiquitous wireless communication networks.
	With multimodal sensing, CT can evolve to predict CSI that closely resembles actual channels. This article outlines recent CT-related activities with different conceptual names. Then, the concept, key features, and two case studies of CT have been discussed. To promote CT deployment in wireless networks, a modularized design architecture of CT has been presented. The architecture includes versatile model configuration of full scenarios, multimodal cooperative sensing, and FL-based lightweight CT training.
	New opportunities for CT in enabling technologies for ubiquitous connectivity and future research challenges have also been discussed.

	\bibliographystyle{IEEEtran}
	\bibliography{ref.bib}

\end{document}